\documentclass[sigconf]{acmart}

\AtBeginDocument{%
  }

\copyrightyear{2023}
\acmYear{2023}
\setcopyright{acmlicensed}\acmConference[KDD '23]{Proceedings of the 29th ACM SIGKDD Conference on Knowledge Discovery and Data Mining}{August 6--10, 2023}{Long Beach, CA, USA}
\acmBooktitle{Proceedings of the 29th ACM SIGKDD Conference on Knowledge Discovery and Data Mining (KDD '23), August 6--10, 2023, Long Beach, CA, USA}
\acmPrice{15.00}
\acmDOI{10.1145/3580305.3599892}
\acmISBN{979-8-4007-0103-0/23/08}

\usepackage{multirow}
\usepackage{mathtools}
\usepackage{bm}
\usepackage{subcaption}
\usepackage[noend]{algorithm2e}
\usepackage{bbold}
\usepackage{balance}
\SetKwInput{KwParameters}{Parameters} 
\SetKw{KwAnd}{and}

\DeclarePairedDelimiter\abs{\lvert}{\rvert}
\newcommand{\bx}{\boldsymbol{x}}
\newcommand{\bz}{\boldsymbol{z}}
\newcommand{\bX}{\boldsymbol{X}}
\newcommand{\bZ}{\boldsymbol{Z}}
\newcommand{\by}{\boldsymbol{y}}
\newcommand{\ymax}{y_\text{max}}
\newcommand{\br}{\boldsymbol{r}}
\newcommand{\rmax}{r_\text{max}}
\newcommand{\bs}{\boldsymbol{s}}
\newcommand{\bd}{\boldsymbol{d}}
\newcommand{\bh}{\boldsymbol{h}}
\newcommand{\bq}{\boldsymbol{q}}
\newcommand{\bv}{\boldsymbol{v}}
\newcommand{\bk}{\boldsymbol{k}}
\newcommand{\bomega}{\boldsymbol{\omega}}
\newcommand{\bsigma}{\boldsymbol{\sigma}}

\newcommand{\SA}{\text{SA}}
\newcommand{\TF}{\text{TF}}
\newcommand{\CLS}{\texttt{[CLS]}}
\newcommand{\concat}{\mathbin\Vert}
\newcommand{\ndcgy}{\text{NDCG}_y}
\newcommand{\ndcgr}{\text{NDCG}_r}
\newcommand{\unif}{\mathcal{U}}

\newtheorem{assumption}{Assumption}

\begin{document}

\title{RankFormer: Listwise Learning-to-Rank Using Listwide Labels}

\author{Maarten Buyl}
\affiliation{%
  \institution{Amazon}
  \city{Luxembourg}
  \country{Luxembourg}
}
\affiliation{%
  \institution{Ghent University}
  \city{Ghent}
  \country{Belgium}
}
\email{maarten.buyl@ugent.be}

\author{Paul Missault}
\affiliation{%
  \institution{Amazon}
  \city{Luxembourg}
  \country{Luxembourg}
}
\email{pmissaul@amazon.com}

\author{Pierre-Antoine Sondag}
\affiliation{%
  \institution{Amazon}
  \city{Luxembourg}
  \country{Luxembourg}
}
\email{pierreas@amazon.com}


\begin{abstract}
Web applications where users are presented with a limited selection of items have long employed ranking models to put the most relevant results first. Any feedback received from users is typically assumed to reflect a relative judgement on the utility of items, e.g. a user clicking on an item only implies it is better than items not clicked in the same ranked list. Hence, the objectives optimized in Learning-to-Rank (LTR) tend to be pairwise or listwise. 

Yet, by only viewing feedback as relative, we neglect the user's absolute feedback on the list's overall quality, e.g. when no items in the selection are clicked. We thus reconsider the standard LTR paradigm and argue the benefits of learning from this listwide signal. To this end, we propose the RankFormer as an architecture that, with a Transformer at its core, can jointly optimize a novel listwide assessment objective and a traditional listwise LTR objective.

We simulate implicit feedback on public datasets and observe that the RankFormer succeeds in benefitting from listwide signals. Additionally, we conduct experiments in e-commerce on Amazon Search data and find the RankFormer to be superior to all baselines offline. An online experiment shows that knowledge distillation can be used to find immediate practical use for the RankFormer.
\end{abstract}



\maketitle
\section{Introduction}
Ranking problems naturally arise in applications where a limited selection of items is presented to a user. A classic example is web search engines \cite{chapelleYahooLearningRank2011}, which show the most relevant URLs to a query. Similarly, e-commerce websites want to show products that customers want to buy \cite{wuTurningClicksPurchases2018} and streaming services show content that they want to watch \cite{gomez-uribeNetflixRecommenderSystem2016}. The user wants to find the item that offers the most utility, as quickly as possible. The selection is thus shown in an ordering, e.g. a list, where the `best' items are shown first. 

The field of Learning-to-Rank (LTR) is well-established in machine learning as a way to form these rankings based on user feedback on previous lists \cite{caoLearningRankPairwise2007, burgesRanknetLambdarankLambdamart2010, liuLearningRankInformation2009}. In many real-world applications, this feedback is implicit: it is collected indirectly from user behavior. For example, if users often purchase a certain product when shown in response to a certain query, then that product should probably be ranked highly in future responses to that query. 

Traditionally, the models behind LTR are simple regressors, such as Gradient-Boosted Decision Trees (GBDTs) \cite{friedmanGreedyFunctionApproximation2001} or neural networks \cite{burgesLearningRankUsing2005}, that compute a \textit{pointwise} score, i.e. a score that is independently determined for every item in a list. Yet it is a fundamental aspect of LTR that any feedback about particular items should be seen in relation to feedback to other items in the list, since user feedback tends to reflect \textit{relative} utility judgments rather than \textit{absolute} judgments \cite{joachimsAccuratelyInterpretingClickthrough2005}. Hence, it is misleading to make \textit{pointwise} comparisons between a ranking model's score for an item and its feedback label. Rather, \textit{pairwise} \cite{burgesRanknetLambdarankLambdamart2010} objectives are used that compare pairs of scores to pairs of labels, or \textit{listwise} \cite{caoLearningRankPairwise2007} objectives that jointly compare scores to labels for the entire list. 

However, a critical shortcoming of pairwise and listwise approaches is that they \textit{only} consider user feedback as a relative judgement. In particular, lists that did not receive any feedback tend to be ignored. Yet this absolute judgement of lists should not be disregarded. In e-commerce, lists that resulted in many purchases are likely to have an overall better selection than lists without any user interaction. In the latter case, the overall selection may be poor for many reasons: because the user gave a bad search query, because the selection of products is too expensive, or because the user is looking for new items that have seen little interaction so far.

\begin{figure*}
    \includegraphics[width=\textwidth]{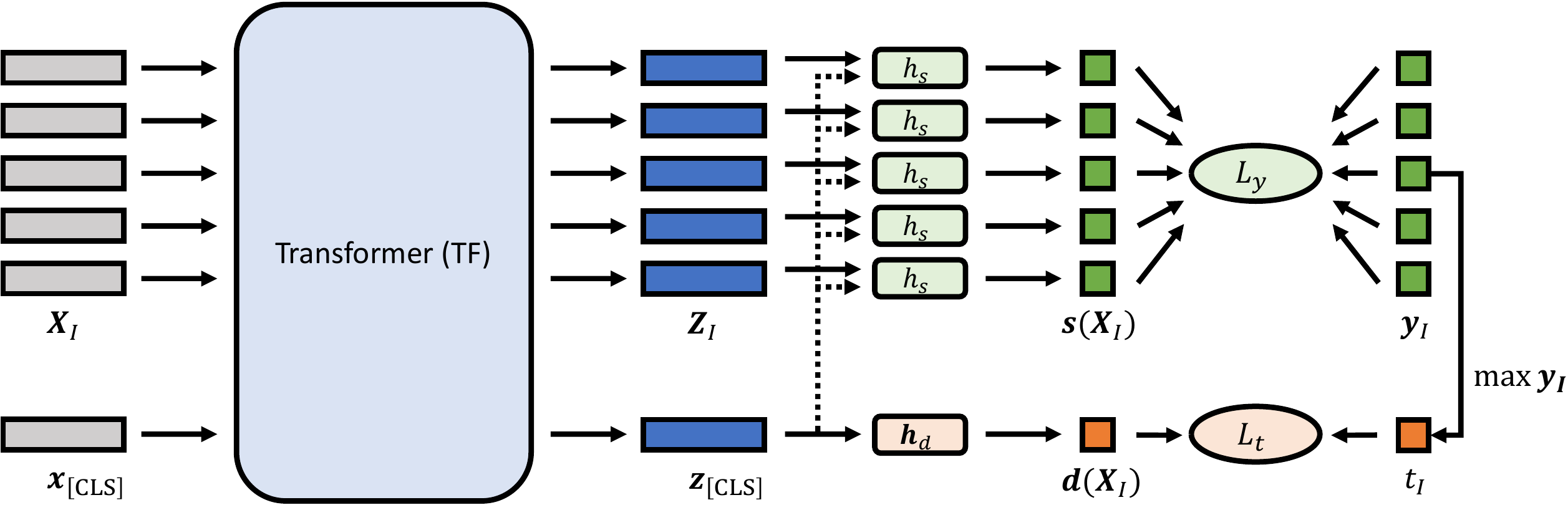}
    \caption{The RankFormer architecture. Features $\bX_I$ of selection $I$, accompanied by the $\CLS$ token vector $\bx_\CLS$, are sent through the listwise Transformer TF. The returned features $\bZ_I$ each correspond with an item $i \in I$, whereas the returned $\bz_\CLS$ vector models the overall list quality. The transformed vectors $\bz_i$ of individual items $i \in I$ are each concatenated with $\bz_\CLS$ and sent through a shared scoring head $h_s$ to obtain scores $\bs(\bX_I)$, which are optimized with a list\textit{wise} LTR loss $L_y$ based on the item feedback labels $\by_I$. Additionally, $\bz_\CLS$ is sent through a separate head $\bh_d$ to generate the list quality prediction $\bd(\bX_I)$, which should match the $t_I$, i.e. the highest item label in the list, and is optimized with a list\textit{wide} loss $L_t$.}\label{fig:rankformer} 
\end{figure*}

\paragraph{Contributions} 
\begin{enumerate}
    \item We make the case for learning from \textit{listwide} labels, i.e. absolute judgments of a list's overall quality, and formalize an approach to derive them as the list's top individual label.
    \item We propose the \textit{RankFormer} architecture, illustrated in Fig.~\ref{fig:rankformer}. At its core, the RankFormer utilizes the \textit{listwise} Transformer \cite{vaswaniAttentionAllYou2017} architecture that was recently applied to perform listwise ranking \cite{pangSetRankLearningPermutationInvariant2020,pobrotynContextAwareLearningRank2021,qinARENEURALRANKERS2021}. It combines this listwise LTR objective with a listwide objective that performs an overall quality assessment of the list and compares it to the listwide labels. Thus, the RankFormer explicitly models the overall list quality and generates individual item ranking scores relative to that overall quality. 
    \item A key use case for learning from listwide objectives is to learn from lists that were presented to users but received no implicit feedback. Yet, such feedback is not provided in the publicly available datasets that are popular in LTR benchmarks, where only \textit{explicit} feedback is given. We therefore simulate implicit feedback on public datasets and observe that the RankFormer successfully leverages its listwide objective to improve upon other neural ranking methods. 
    \item We conduct experiments on data collected from Amazon Search in both an offline and online setting. Offline, the RankFormer is superior to all baselines. Online, this advantage can be maintained when distilling the RankFormer to a simple GBDT model. Despite its complexity, the RankFormer is thus readily put into practice.
\end{enumerate}

\section{Related Work}
Learning-to-Rank (LTR) provides tools, e.g. objective functions, to learn from feedback labels such that a ranking model is improved. To this end, \textit{neural} ranking uses neural networks as the model.

A distinction is made with (neural) \textit{Information Retrieval} \cite{guoDeepLookNeural2020}, which often applies (neural) ranking to obtain information related to a query from a single collection of information resources. Though that field aims to find the \textit{most} relevant resources, it has less of a focus on the LTR objectives themselves and more so on the relevance of individual information resources. Instead, we assume that a limited selection of relevant items has already been made, and our model should now put the most relevant of these on top.

Traditional LTR \cite{liuLearningRankInformation2009} employs a \textit{pointwise} scoring function, i.e. one that computes an individual score for each item independently, and only then compares those scores in pairwise \cite{burgesRanknetLambdarankLambdamart2010} or listwise \cite{caoLearningRankPairwise2007} loss functions. Yet the actual utility of items may depend on the rest of the list, e.g. an item may be much more desireable if it is the cheapest in the selection. Methods have therefore been proposed to compute \textit{listwise} scores directly, i.e. scores that depend on the features of all items in the list. To this end, Ai et al. \cite{aiLearningDeepListwise2018a} proposed the Deep Listwise Context Model, which uses a recurrent neural network to take other inputs into account when reranking the top documents in information retrieval. This approach was further formalized to general multivariate scoring functions \cite{aiLearningGroupwiseMultivariate2019}.  

A list of items can be seen as a sequence, hence many approaches from Natural Language Processing (NLP) can readily be applied to ranking. Most notably, the Transformer \cite{vaswaniAttentionAllYou2017} architecture has been used to learn from listwise ranking contexts using self-attention, which we refer to as \textit{listwise} Transformers \cite{peiPersonalizedRerankingRecommendation2019,pasumarthiPermutationEquivariantDocument2020,pangSetRankLearningPermutationInvariant2020,qinARENEURALRANKERS2021,pobrotynContextAwareLearningRank2021,zhuangCrossPositionalAttentionDebiasing2021,liPEARPersonalizedReranking2022}. As a sequence-to-sequence architecture, the Transformer here considers sequences of items instead of words. This form of \textit{listwise} attention is thus orthogonal to any usage Transformers that may be used to compute the feature vectors of individual items.

Particular to our approach, the RankFormer, is that we use the listwise Transformer to combine the listwise LTR objective with a \textit{listwide} objective, for which the overall quality of the list is explicitly modelled. The Deep Listwise Context Model proposed by Ai et al. \cite{aiLearningDeepListwise2018a} also models the list context with a recurrent neural network, yet this model is not used to actually predict listwide signals. The most related work to ours is the Personalized Re-ranking with Contextualized Transformer for Recommendation (PEAR), recently proposed by Li et al. \cite{liPEARPersonalizedReranking2022}. In the personalized recommendation setting, the PEAR model combines a pointwise ranking objective, i.e. predicting whether the item is clicked, with a listwide objective where it should be predicted whether \textit{any} item is clicked. Our work formalizes the listwide objective as an ordinal loss and combines it with a listwise ranking objective in the general LTR context.

In recent work on score calibration in LTR, Yan et al. \cite{yanScaleCalibrationDeep2022} aim to predict scores that are calibrated to the item labels, as their scores are both used for ranking \textit{and} for downstream tasks. Though our proposed model does not require scores to be calibrated, we leverage their observation that popular pair- and listwise loss functions are invariant to the absolute scale of scores, which causes them to miss an important signal during training.

Finally, note that much of the recent LTR literature has been interested in \textit{unbiased} LTR \cite{joachimsUnbiasedLearningtoRankBiased2017}, which considers the bias in clicking behavior depending on the order in which items are shown to the user. We will not consider this bias in our proposed model to reduce its complexity, i.e. our proposed model is permutation-invariant with respect to the list ordering. A study on how the overall quality of the list (and thus also the listwide label) may depend on the list ordering is left to future work.


\section{Background}\label{sec:background}
Our contributions are aimed at the ranking problem, which we formalize in Sec.~\ref{sec:ranking}. In Sec.~\ref{sec:ltr}, we briefly review how such problems are typically solved through Learning-to-Rank. A particular ranking model that is core to our proposed method is the listwise Transformer, which we discuss in Sec.~\ref{sec:transformer}.

\subsection{Ranking}\label{sec:ranking}
We use $\bx_i \in \mathbb{R}^{d_x}$ to denote the real-valued, $d_x$-dimensional feature vector corresponding with the occurrence item $i \in I$ in a selection list $I$. For example in search engines, these features may be solely based on the item itself or also dependant on a search query and other contextual information. Based on the response to the selection, feedback values $y_i \in \mathcal{Y} = \{0, ..., \ymax\}$ are gathered for each item $i \in I$. The feedback values are assumed to be ordinal, where higher values indicate that the user had a stronger preference for the item. In this work, we focus on \textit{implicit} feedback, e.g. in the form of clicks and purchases, which is inferred from a user's overall interaction with the selection. This is opposed to \textit{explicit} feedback, where the user directly assign scores to items. A key difference is that implicit labels often make no distinction between `negative' and `missing' feedback. For example, $y_i = 0$ may indicate that item $i$ was bad, or that item $i$ was ignored and thus never explicitly considered.

Let $\bX_I = (\bx_i^\top)_{i \in I}$ denote the matrix of all feature vectors in list $I$ and $\by_I = (y_i)_{i \in I}$ the vector of all its labels. The ranking dataset $\mathcal{D} = \{(\bX_{I}, \by_{I}) \mid I \in \mathcal{I}\}$ then gathers all lists in the superset $\mathcal{I}$. Note that in realistic search engine settings, it is likely that the same item shows up in many lists. As such, when we refer to item $i$, we mean the occurrence of an item in list $I$ where it had features $\bx_i$ and label $y_i$. Moreover, our experiments allow for the case where the same selection with the same features shows up in lists $A$ and $B$, though the collected user feedback may be different: $\bX_A = \bX_B \centernot \implies \by_A = \by_B$.

\subsection{Learning-to-Rank}\label{sec:ltr}
The goal of Learning-to-Rank (LTR) \cite{liuLearningRankInformation2009} is to produce rankings that lead to a high utility. Typically, this is done using a scoring function $\bs: \mathbb{R}^{\abs{I} \times d_x} \to \mathbb{R}^{\abs{I}}$ that produces a vector of scores $\bs(\bX_I)$ for list $I$ with features $\bX_I$. The individual items $i$ are then ranked in the descending order of $\bs(\bX_I)_i$ scores. Traditional scoring functions are \textit{pointwise}, i.e. each score $\bs(\bX_I)_i = s(\bx_i)$ is computed individually per feature vector $\bx_i$.

In \textit{Neural} LTR approaches, $\bs$ is a parameterized neural network for which a loss function is optimized that compares scores $\bs(\bX_I)$ to labels $\by_I$. This comparison is either made \textit{pointwise} \cite{liMcRankLearningRank2007} (i.e. $\bs(\bX_I)_i$ with $y_i$), \textit{pairwise} \cite{burgesLearningRankUsing2005, burgesLearningRankNonsmooth2006} (i.e. $(\bs(\bX_I)_i, \bs(\bX_I)_j)$ with $(y_i,y_j)$ for $i,j \in I$), or \textit{listwise} (i.e. $\bs(\bX_I)$ with $\by_I$). In particular, we use the \textit{ListNet} \cite{caoLearningRankPairwise2007} or \textit{Softmax} loss. The Softmax loss $L_y$ is defined as 
\begin{equation}\label{eq:listnet}
    L_y(\bs(\bX_I), \by_I) = - \by_I^\top \log \bsigma(\bs(\bX_I))
\end{equation}
which is simply the dot-product of the label vector $\by_I$ and the elementwise log of the softmax function $\bsigma(\mathbf{a})_i = \frac{\exp(\mathbf{a}_i)}{\sum_{j \in I} \exp(\mathbf{a}_j)}$ for $i \in I$, applied to the scores $\bs(\bX_I)$.

If the ranking is done according to a specific stochastic model (i.e. ranking permutations are randomly sampled from a distribution that is influenced by the scores $s(\bX)$) and if the label domain is binary (i.e. $\forall i: y_i \in \{0, 1\}$), then Softmax loss indicates the likelihood of item $i$ to be ranked first \cite{caoLearningRankPairwise2007,bruchAnalysisSoftmaxCross2019}. Though we do not constrain the label domain to be binary, we have found the Softmax loss to show strong empirical results while being simple and efficient. This is aligned with recent findings in similar work \cite{qinARENEURALRANKERS2021,pobrotynContextAwareLearningRank2021}.

\subsection{The Listwise Transformer}\label{sec:transformer}
Though scoring functions $\bs$ usually compute pointwise scores, i.e. independently per item, it may be beneficial to make the score of an item $i$ dependant on other items $\{j \mid j \in I \wedge j \neq i\}$ in the selection $I$. E.g. in e-commerce, if a product $i$ is much cheaper than the other products that will be shown, then price-related features in $\bx_i$ may be more important than when all products have the same price. 

A \textit{listwise} scoring function $\bs$ can be constructed by having a computation step that transforms the feature matrix $\bX_I$ in each list $I$ such that information is shared between items $i \in I$. This kind of transformation can be learned end-to-end by applying an attention mechanism over $\bX_I$, e.g. through the use of the \textit{Transformer} \cite{vaswaniAttentionAllYou2017} architecture that has seen tremendous success in Natural Language Processing (NLP) and has recently also been applied to ranking problems \cite{pangSetRankLearningPermutationInvariant2020,qinARENEURALRANKERS2021,pobrotynContextAwareLearningRank2021,liPEARPersonalizedReranking2022}.

The core component of the Transformer architecture is the \textit{Self-Attention} (SA) layer. Here, each feature vector $\bx_i \in \mathbb{R}^{d_x}$ in the feature matrix $\bX_I \in \mathbb{R}^{\abs{I} \times d_x}$ is independently sent through a shared \textit{query} function $\bq: \mathbb{R}^{d_x} \to \mathbb{R}^{d_{\text{att}}}$, \textit{key} function $\bk: \mathbb{R}^{d_x} \to \mathbb{R}^{d_{\text{att}}}$, and \textit{value} function $\bv: \mathbb{R}^{d_\text{val}} \to \mathbb{R}^{d_{\text{val}}}$, with $d_{\text{att}} = d_{\text{val}} = d_x$ in our implementation. For each input vector $\bx_i$, the output value of the SA layer is then the listwise sum of value matrix $\bv(\bX_I) = (\bv(\bx_i)^\top)_{i \in I}$, weighted by the softmax $\bsigma$ of the dot-product between its query vector $\bq(\bx_i)$ and the matrix of key vectors $\bk(\bX_I) = (\bk(\bx_i)^\top)_{i \in I}$. The SA operation is thus defined as
\begin{equation}
    \SA(\bX_I)_i = \bv(\bX_I)^\top \bsigma \left(\frac{\bk(\bX_I) \bq(\bx_i)}{\sqrt{d_\text{att}}}\right).
\end{equation} 

A typical Transformer encoder layer is made up of an SA block followed by a shared \textit{Feed-Forward} (FF) network that processes each vector in the sequence independently. Furthermore, the SA and FF blocks often contain residual, Dropout and LayerNorm layers. For a full discussion, we refer to the original work \cite{vaswaniAttentionAllYou2017} and comparative technical overviews \cite{phuongFormalAlgorithmsTransformers2022}. Details on our exact implementation can be found in Appendix \ref{app:transformer_details}.

We define $\TF: \mathbb{R}^{\abs{I} \times d_x} \to \mathbb{R}^{\abs{I} \times d_x}$ as the full Transformer encoder that, given feature matrix $\bX_I$ as input, returns the list-aware feature matrix $\bZ_I = \TF(\bX_I)$. The transformed feature vectors $\bz_i$ correspond with items $i$, but were computed knowing the context of other items in list $I$. To finally construct the \textit{listwise} scoring function $\bs$, we therefore apply a pointwise, shared scoring head $h_s: \mathbb{R}^{d_x} \to \mathbb{R}$ to each $z_i$:
\begin{equation}
    \bs(\bX_I)_i = h_s(\TF(\bX_I)_i)
\end{equation}
where $h_s$ is a learnable, feed-forward network.

\section{Listwide Labels}\label{sec:listwide}
In contrast to the \textit{listwise} LTR defined in Sec.~\ref{sec:ltr}, we now discuss \textit{listwide} labels. These quantify the overall quality of the list. 

A key use case of listwide labels is the ability to learn from lists where all labels are zero, which we motivate in Sec.~\ref{sec:constant_labels}. We then formulate a precise definition of listwide labels in Sec.~\ref{sec:deriving} and a strategy to learn from them in Sec.~\ref{sec:learning_listwide}.

\subsection{Lists without Relevancy Labels}\label{sec:constant_labels}
Performing listwise LTR can deliver strong results because it jointly evaluates the scoring function $\bs(\bX_I)$ over all items $i \in I$ with respect to the user feedback labels $\by_I$. Yet, listwise (and pairwise) LTR loss functions only consider the values of scores and labels at separate, relative scales per list. For example, if a label $y_i$ is low, then it will only be considered low in that list by pair- and listwise loss functions if there are other labels $y_j > y_i$ in the same list. Many such loss functions are even completely invariant to global translations of the scores vector $\bs(\bX_I)$, which means scores are not necessarily calibrated to the labels they try to align with \cite{yanScaleCalibrationDeep2022}.

Hence, constant lists with value $c$, i.e. $\forall i \in I: y_i = c$, provide no useful information to a pair- or listwise loss function \cite{wangLearningRankSelection2016}. Indeed, for constant lists with $c \neq 0$ the $L_y$ loss in Eq.~\ref{eq:listnet} will push all scores $\bs(\bX_I)$ to simply be constant, regardless of whether $c$ is high or low. In the particular case of lists without relevancy labels, i.e. $\forall i \in I: y_i = 0$, we even have the Softmax loss $L_y(\cdot, \mathbf{0}) = 0$. Though pointwise loss functions \textit{do} consider the absolute scale of individual labels $y_i$, they are otherwise empirically inferior to the best pair- and listwise loss functions in LTR benchmarks \cite{qinLETORBenchmarkCollection2010,qinARENEURALRANKERS2021}. 

In practice, lists without relevancy labels therefore tend to be ignored during optimization. This either happens \textit{implicitly} because pair- or listwise loss functions are typically zero or constant with respect to the scores, or \textit{explicitly} because they are filtered out during preprocessing \cite{wangLearningRankSelection2016,bruchAnalysisSoftmaxCross2019,pobrotynContextAwareLearningRank2021}. 

However, lists without relevancy labels are actually common in real-world implicit feedback, since users often do not interact with lists. We argue such lists should not be ignored, as this could be an important signal about the quality of the ranking. For example in the e-commerce setting, $\by_I = \mathbf{0}$ could mean that none of the products in $I$ where desirable, so the user may have tried a different query or store. Indeed, internal data at Amazon Search shows that many queries do not result in any user feedback. To better capture this signal, we propose that the score function $\bs$ should learn that the products in the selection all had the same (poor) quality.

A straightforward solution could therefore be to exclusively use pointwise LTR loss functions for those lists. Yet, this approach still misses that the constant quality of label vectors such as $\by_I = \mathbf{0}$ may, in part, be influenced by the \textit{overall} quality of $I$. As concluded in the user study by Joachims et al. \cite{joachimsAccuratelyInterpretingClickthrough2005}: "clickthrough data does not convey \textit{absolute} relevance judgments, but partial \textit{relative} relevance judgments for the links the user evaluated". 

Going one step further, we suggest that it may not suffice to treat lists with $\by_I = \mathbf{0}$ as a combination of items that \textit{individually} had the same poor quality, but rather as a set of items that \textit{together} received no feedback. Hence, we propose to learn from this signal at the list level, as opposed to decomposing it over the item level.

\subsection{Deriving Listwide Labels}\label{sec:deriving}
We introduce the notation $t_I \in \mathcal{T}$ to refer to the overall quality of the list $I$, i.e. the \textit{listwide} label. To generate informative listwide labels, in particular for $\by_I = \mathbf{0}$, we derive $t_L$ from the item labels. In our case, we set $t_I$ as the \textit{maximal} implicit feedback label:
\begin{equation}\label{eq:t_def}
    t_I = \max_{i\in I} y_i
\end{equation}
where it is practical that $t_I = 0 \iff \by_I = \mathbf{0}$ with item labels $y_i \in \mathcal{Y} =  \{0, ..., \ymax\}$. Also, note that the domain of the item labels is inherited by the listwide labels $t_I \in \mathcal{T} = \mathcal{Y}$.

This formulation for $t_I$ is motivated by the observation that in most ranking contexts, users are typically looking for a single `thing'. Overall user satisfaction, which is a long-term goal in web and e-commerce search, can thus be approximated by observing whether the user's overall goal was reached. For example, that goal might be "finding a relevant document" during a web search or "buying a suitable product" in e-commerce.

\subsection{Learning from Listwide Labels}\label{sec:learning_listwide}
To learn from listwide labels, we define $d: \mathbb{R}^{\abs{I} \times d_x} \to \mathcal{T}$ as the function that estimates the listwide quality $t_I$. For a parameterized $d$, we can then optimize the listwide loss function $L_t(d(\bX_I), t_I)$ for list $I$. As the values $d(\bX_I)$ and $t_I$ are singletons with a domain inherited from the item labels (through Eq.~\ref{eq:t_def}), any pointwise loss function for the item labels in LTR is also applicable for the definition of $L_t$. 

Considering that $\mathcal{T} = \mathcal{Y} = \{0, ..., \ymax\}$ 
is ordinal, we make use of the \textit{Ordinal} loss function \cite{niuOrdinalRegressionMultiple2016} as applied to ranking \cite{pobrotynContextAwareLearningRank2021}. To this end, we use $\bomega: \mathcal{T} \to \{0, 1\}^{\ymax}$ as the function that converts the listwide labels $t_I$ into their ordinal, one-hot encoding $\bomega(t_I)$, i.e.
\begin{equation}
    \bomega(t_I)_k = 
    \begin{cases}
        1 & \text{ if } t_I \geq k\\
        0 & \text{ else}
    \end{cases}
\end{equation}
with $k \in \{1, ..., \ymax\}$.

For example with $\mathcal{T} = \{0, 1, 2\}$, we would have vectors $\bomega(0) = (0, 0)$, $\bomega(1) = (1, 0)$ and $\bomega(2) = (1, 1)$.

Furthermore, we now change the listwide predictor such that it is also vector-valued, i.e. $\bd: \mathbb{R}^{\abs{I} \times d_x} \to [0, 1]^{\ymax}$
The \textit{Ordinal} loss function $L_t$ is then defined as
\begin{equation}
    L_t(\bd(\bX_I), t_I) = \sum_{k = 1}^{\ymax} \text{BCE}\left(\bd(\bX_I)_k, \bomega(t_I)_k\right)
\end{equation}
which is the sum of the Binary Cross-Entropy (BCE) losses for each non-zero $k \in \mathcal{T}$. The ordinality of $L_t$ results from the fact that if $\bd(\bX_I)_k$ is high, then the estimated likelihood of $\bd(\bX_I)_l$ for $l < k$ should also be high. For example in e-commerce, if $t_I = 1$ implies that the user clicked something and $t_I = 2$ implies they purchased something, then $\bd$ should only predict that a purchase was made if it is also predicting a click.

Finally, note that the benefit of listwide assessments not only results from being able to learn from lists with constant feedback (e.g. $\by_I = \mathbf{0}$). It also has the practical advantage that it allows us to make a quality assessment about selections that is easy to aggregate over large sets of selections. For example in e-commerce, predicting $t_I$ allows us to judge whether new selections $I$ will be successful or not, before they are shown to customers. 

\section{RankFormer}\label{sec:rankformer}
Learning from the list\textit{wide} labels discussed in Sec.~\ref{sec:listwide} allows us to fill in the gaps where list\textit{wise} LTR, as discussed in Sec.~\ref{sec:ltr}, is not possible (e.g. when $\by_I = \mathbf{0}$). We therefore propose the \textit{RankFormer} architecture, illustrated in Fig.~\ref{fig:rankformer}, which is an extension of the listwise Transformer architecture discussed in Sec.~\ref{sec:transformer} that performs both listwise ranking and listwide quality predictions. By explicitly assessing the overall quality of a selection $I$, the individual scores for items $i \in I$ can be computed in the context of that overall quality.

To make predictions about the entire lists, we follow the standard NLP approach by adding a shared $\CLS$ token to each list. Since all tokens in listwise Transformers are expected to be feature vectors, we also define the learnable parameter vector $\bx_\CLS$ that is the same for every list. The items in the appended list $I^* = I \cup \CLS$, with feature matrix $\bX_{I^*}$, are then provided as input to the Transformer operation $\TF$ as normal. From the transformed features $\bZ_{I^*} = \TF(\bX_{I^*})$ returned as output, we extract the $\bz_\CLS$ vector that should reflect the information in $\bX_I$ that was particularly relevant to judge the overall quality of $I$. 

The actual listwide prediction $\bd$ is made by applying a separate feed-forward network block  $\bh_d: \mathbb{R}^{d_x} \to [0, 1]^{\ymax}$ to $\bz_\CLS$:
\begin{equation}
    \bd(\bX_I) = \bh_d(\TF(\bX_{I^*})_\CLS)
\end{equation}
where we recall that $\bh_d$ is vector-valued such that the Ordinal loss can be computed.

The listwise scores $\bs(\bX_I)$ could be computed as normal. However, we can make sure that the individual scores assigned to transformed feature vectors $\bz_i$ for $i \in I$ are benefitting from the overall list quality information $\bz_\CLS$ by concatenating the latter with each of the $\bz_i$ vectors. The listwise LTR scores $\bs(\bX_I)$ are thus computed as
\begin{equation}
    \forall i \in I: \bs(\bX_I)_i = h_s(\TF(\bX_{I^*})_i \concat \TF(\bX_{I^*})_\CLS)
\end{equation}
with $\concat$ the concatenation operator and with the shared scoring head now $h_s: \mathbb{R}^{2d_x} \to \mathbb{R}$.

The RankFormer, with its learnable parameters in $\bx_\CLS$, $\TF$, $h_s$ and $\bh_d$, thus contains all the necessary components to perform listwise LTR while making listwide predictions. A straightforward way to optimize these parameters is through the minimization of the sum of the listwise loss $L_y$ and the listwide loss $L_t$:
\begin{align}
    \min_{\bx_\CLS, \TF, h_s, \bh_d} \sum_{I \in \mathcal{I}} L_y(s(\bX_I), \by_I) + \alpha L_t(\bd(\bX_I), t_I).
\end{align}
where the hyperparameter $\alpha$ controls the strength of $L_t$.

Ending our discussion of the RankFormer, we note that it does not use the positional embeddings that are popular in the application of Transformers in NLP \cite{vaswaniAttentionAllYou2017}. The listwise LTR scores are thus permutation-equivariant \cite{pangSetRankLearningPermutationInvariant2020} with respect to $\bX$, whereas the listwide quality predictions are permutation-invariant. Here, the use of positional embeddings is not necessary because we assume the selection $I$ to be unordered with no prior ranking. We refer to related work \cite{pobrotynContextAwareLearningRank2021} for a study on the impact of this assumption.   

\section{Experiments on Public Datasets}\label{sec:public}
We performed experiments on three public datasets: MSLR-\textsc{WEB30k} \cite{qinIntroducingLETORDatasets2013}, \textsc{Istella}-S \cite{lucchesePostLearningOptimizationTree2016} and Set 1 of the \textsc{Yahoo} \cite{chapelleYahooLearningRank2011} dataset. All consist of web search data, where humans were asked to assign an explicit relevance label to each URL in the selection.

However, our focus is on learning from implicit labels, which are far more abundant in realistic applications. As is done in Unbiased LTR literature \cite{joachimsUnbiasedLearningtoRankBiased2017}, we therefore simulate implicit labels in a manner that aligns with realistic user behavior seen in industry. This is discussed in Sec.~\ref{sec:simulating}.

We further detail our evaluation setup in Sec.~\ref{sec:eval} and finally report our results in Sec.~\ref{sec:results}.

\subsection{Simulating Implicit Feedback}\label{sec:simulating}
Let $r_i \in \{0, ..., \rmax\}$ denote the \textit{explicit} relevance label assigned to item $i \in I$. It so happens that all our datasets have $\rmax = 4$. For example in \textsc{Yahoo}, gradings $r_i$ correspond with
\begin{equation*}
    \{\texttt{bad}, \texttt{fair}, \texttt{good}, \texttt{excellent}, \texttt{perfect}\}.
\end{equation*}

Recall that \textit{implicit} relevance labels $y_i \in \{0, ..., \ymax\}$ are inferred from user behavior. In what follows, we explicitly set $\ymax = 2$ and interpret the domain of $y_i$ as 
\begin{equation*}
    \{\texttt{seen}, \texttt{click}, \texttt{conversion}\}.
\end{equation*}

\begin{assumption}\label{ass:correlation}
Implicit labels $y_i$ are positively correlated with explicit labels $r_i$.
\end{assumption}

To put Assumption~\ref{ass:correlation} into practice, we use the function $\rho$, which maps the relevance grades $r_i$ to a probability of relevance \cite{chapelleExpectedReciprocalRank2009,aiUnbiasedLearningRank2018}:
\begin{equation}
    \rho(r_i) = \frac{2^{r_i} - 1}{2^{\rmax} - 1}.
\end{equation}

In an effort to align the exact relationship between $r_i$ and $y_i$ with experiences from industry, we perform our simulation in three stages: \textit{Selection} where subsets of lists are sampled, \textit{Intent} where an overall list assessment is made, and \textit{Interaction} where the actual implicit label $y_i$ is generated.

\subsubsection{Selection}\label{sec:select}
A common effect in LTR applications is \textit{selection bias} \cite{jagermanModelInterveneComparison2019}, i.e. though a selection $J \in \mathcal{I}$ was gathered, users often only examine a subset of the selection $I \subset J$ with a maximal length $N_s$.

\begin{assumption}\label{ass:select}
    The actual selections $I$ seen by users are small.
\end{assumption}

For example, Assumption~\ref{ass:select} holds when only $N_s$ items are shown on a single webpage and users do not navigate to later pages. A practical way to generate such lists is to randomly sample up to $N_s$ items without replacement. 

However, this step leads to many sampled lists $I$ that exclusively contain items with low $r_i$ values. In practice, it is indeed realistic that many selections have an overall low relevance, e.g. because users gave an uninformative query. Yet, it may be too difficult to achieve meaningful validation and test set performance on public datasets because many items in the original training set will be removed. For each of the original lists $J$, we therefore perform bootstrapping by sampling $N_b$ selections $I$. This also mimics a scenario where not all users see exactly the same list for the same query

In our experiments, we set $N_s = 16$ and $N_b = 10$.

\begin{figure*}[ht]
    \includegraphics[width=\textwidth]{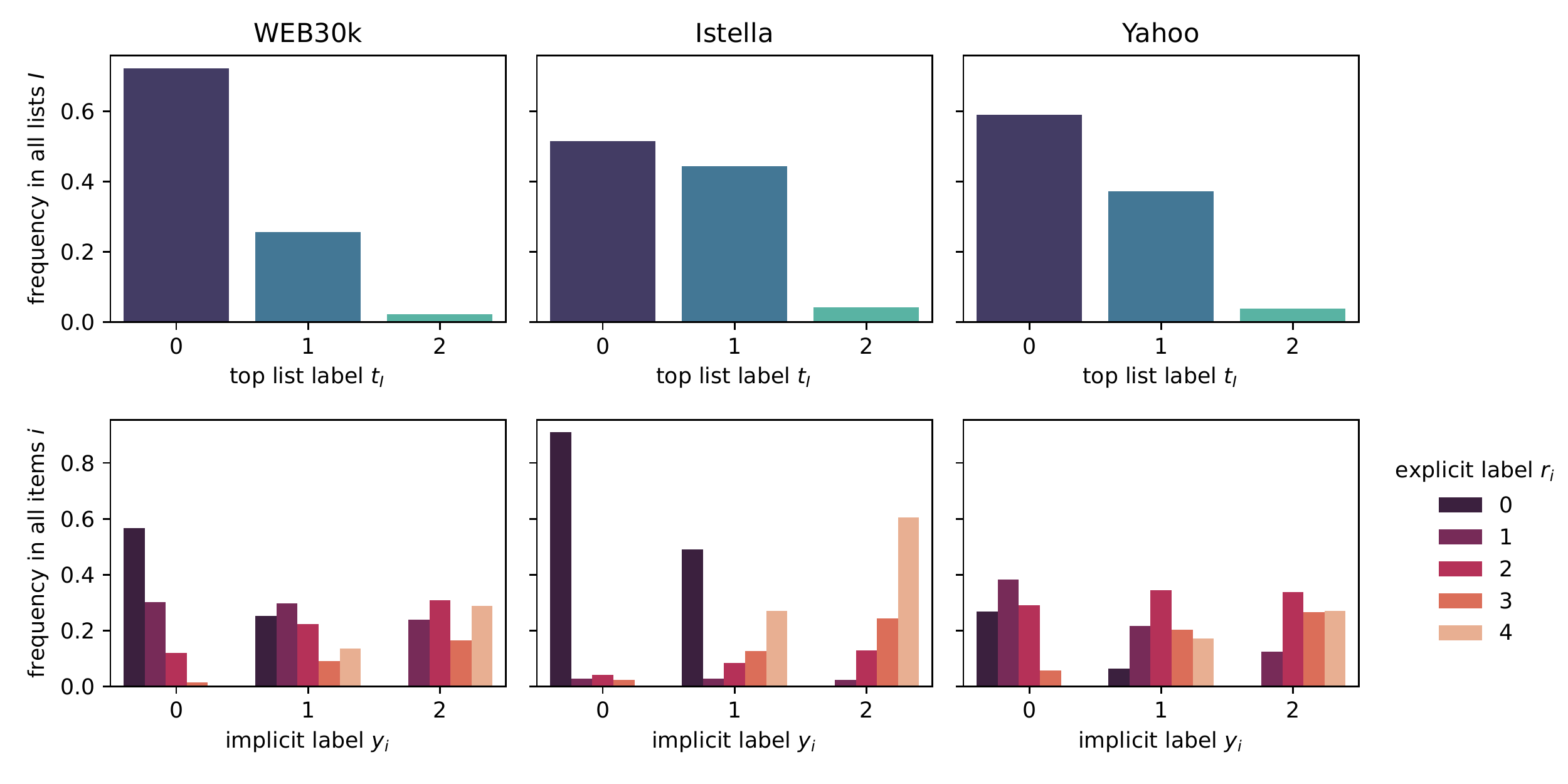}
    \caption{The simulated distribution of the lists' top implicit labels $t_I = \max_{i \in I} y_i$ and the distributions of the individual explicit labels $r_i$ per simulated implicit label $y_i$.}
    \label{fig:labels}
\end{figure*}

\subsubsection{Intent}
Our aim is to explicitly simulate the \textit{intent} of the user in interacting with the list, e.g. whether the user wants to only view the selection, or whether they want to invest more time and also click and buy things \cite{moeBuyingSearchingBrowsing2003}. To this end, we take inspiration from the user study by Liu et al. \cite{liuSkimmingReadingTwostage2014} where it is observed that users first skim many of the items in the list, before giving any of them serious consideration. We assume that the overall relevancy $\br_I = (r_i)_{i \in I}$ observed during this skimming step affects the user intent.

\begin{assumption}\label{ass:overall}
    Users intend stronger interactions with lists $I$ that have an overall higher relevance $\br_{I}$.
\end{assumption}

Let $T_I$ denote the intent of the user for list $I$, which we consider as the maximal action that the user is willing take. Thus $T_I \in \mathcal{Y}$, since we measure user behavior as implicit feedback. To model Assumption~\ref{ass:overall}, we echo the definition of listwide labels in Eq.~\ref*{eq:t_def} and consider the overall list quality as the highest explicit label: 
\begin{equation}\label{eq:intent}
    P(T_I \mid \br_I) = 
    \begin{cases}
        0 & 1 - \rho(\max_{i \in I} r_i)\\
        1 & (1 - \kappa) \rho(\max_{i \in I} r_i)\\
        2 & \kappa \rho(\max_{i \in I} r_i)
\end{cases}
\end{equation}
with $\kappa \in [0, 1]$ a hyperparameter that controls how often clicks result in conversions. It is common that $\kappa \ll 0.5$ in e-commerce settings, since users mostly only want to browse \cite{moeBuyingSearchingBrowsing2003}. 

The pragmatic benefit of Eq.~\ref{eq:intent} is that it allows us to simulate all-zero lists, as they are guaranteed for intent $T_I = 0$. Some simulations of implicit feedback in prior work, such as by Jagerman et al. \cite{jagermanModelInterveneComparison2019}, were also capable of generating all-zero lists. However, they do not do so by simulating the overall intent of users and do not assume a dependency on the overall quality of the list.

In our experiments, we set $\kappa = 0.1$.

\subsubsection{Interaction}
Finally, given the intent $T_I$ for the observed list $I$, we sample individual implicit labels $y_i$.

\begin{assumption}\label{ass:noise}
Higher implicit labels $y_i$ are less noisy.
\end{assumption}

We also motivate Assumption~\ref{ass:noise} from the e-commerce setting, where conversions are more robust than clicks \cite{karmakersantuApplicationLearningRank2017}. For example, this is because higher $y_i$ labels involve a higher (financial) cost, so more thought is put into the action.

For $y_i \in \{0, 1, 2\}$, we first decide conversions ($y_i = 2$) and then clicks ($y_i = 1$) for all items that did not have a conversion, i.e. $y_i < 2$. Given the overall list intent $T_I$, we sample these decisions with likelihoods
\begin{align}
\begin{split}\label{eq:interaction}
    &P(y_i = 2 \mid r_i, T_I = 2) = \rho(r_i)\\
    &P(y_i = 1 \mid r_i, T_I \geq 1, y_i < 2) = \epsilon + (1 - \epsilon) \rho(r_i).  
\end{split}
\end{align}

Here, $\epsilon \in [0, 1]$ is a parameter that represents the noise in click behavior \cite{aiUnbiasedLearningRank2018}. Following Assumption~\ref{ass:noise}, no explicit noise is simulated for $y_i=2$. Moreover, following our definition of the list intent $T_I$, we clarify that $\forall k: P(y_i = k \mid T_I<k) = 0$. 

In experiments, we set $\epsilon = 0.1$ as was previously done in work on unbiased LTR \cite{aiUnbiasedLearningRank2018}. However, in contrast to unbiased LTR settings \cite{aiUnbiasedLearningRank2018, joachimsUnbiasedLearningtoRankBiased2017}, we do not assume that the likelihood of clicks is dependant on the item's position in the shown list. 

\subsection{Evaluation Setup}\label{sec:eval}

\begin{table}[htb]
\caption{Statistics of the public datasets. The selection stage in the simulation keeps at most $N_s = 16$ items per list. Each query is sampled $N_b = 10$ times per original list. For \textsc{WEB30k}, the statistics are computed for Fold 1.}
\label{tab:stats}
\begin{tabular}{cl|ccc}
\multicolumn{2}{c|}{} & \textsc{WEB30k} & \textsc{Istella} & \textsc{Yahoo}\\ \hline\hline
\multicolumn{2}{c|}{\# features} & 136 & 220 & 700 \\\hline
\multirow{3}{*}{\# original lists} & train & 18,919 & 19,245 & 19,944 \\
 & valid & 6,306 & 7,211 & 2,994 \\
 & test & 6,306 & 6,562 & 6,983 \\\hline
 \multirow{2}{*}{avg. length} & original & 119.6 & 103.2 & 23.7 \\
 & sampled & 15.7 & 16.0 & 12.7
\end{tabular}
\end{table}

\begin{table*}[ht]
    \caption{NDCG@10 on the \textsc{WEB30k}, \textsc{Istella} and \textsc{Yahoo} datasets. $\ndcgy$ refers to NDCG as measured on simulated, \textit{implicit} labels. $\ndcgr$ is computed over the original, \textit{explicit} labels. Each result is the mean out of five runs, $\pm$ the standard error. The GBDT with $10^4$ trees is superior in almost all cases, so we show the \textit{two} configurations with the best results in bold.}
    \label{tab:main}
    \begin{tabular}{ll||cc|cc|cc}
    & & \multicolumn{2}{c|}{\textsc{WEB30k}} & \multicolumn{2}{c|}{\textsc{Istella}} & \multicolumn{2}{c}{\textsc{Yahoo}} \\ 
    & & $\ndcgy$ & $\ndcgr$ & $\ndcgy$ & $\ndcgr$ & $\ndcgy$ & $\ndcgr$ \\ \hline \hline
    \multirow{2}{*}{GBDT} & $\text{\# trees} = 10^3$ & $51.17\pm0.17$ & $63.01\pm0.09$ & $67.50\pm0.05$ & $86.99\pm0.02$ & $\bm{66.30\pm0.02}$ & $79.35\pm0.03$\\
    & $\text{\# trees} = 10^4$ & $\bm{51.53\pm0.14}$ & $63.52\pm0.09$ & $\bm{68.42\pm0.04}$ & $\bm{88.75\pm0.01}$ & $\bm{66.82\pm0.04}$ & $\bm{80.71\pm0.02}$ \\ \hline
MLP &  & $50.74\pm0.10$ & $63.13\pm0.07$ & $67.63\pm0.06$ & $87.38\pm0.04$ & $65.94\pm0.08$ & $\bm{79.59\pm0.03}$\\ \hline
\multirow{3}{*}{RankFormer} & $\alpha = 0$ & $51.19\pm0.08$ & $63.78\pm0.10$ & $67.83\pm0.08$ & $87.75\pm0.03$ & $65.82\pm0.03$ & $79.27\pm0.15$\\
    & $\alpha = 0.25$ & $\bm{51.32\pm0.12}$ & $\bm{63.97\pm0.13}$ & $\bm{67.95\pm0.07}$ & $\bm{87.88\pm0.04}$ & $65.95\pm0.09$ & $79.51\pm0.11$\\
    & $\alpha = 1$ & $51.19\pm0.09$ & $\bm{63.84\pm0.08}$ & $67.85\pm0.06$ & $87.82\pm0.02$ & $65.23\pm0.10$ & $78.22\pm0.27$\\
\end{tabular}
\end{table*}

Each configuration is run for five random seeds. For the \textsc{WEB30k} dataset, each seed indexes a different fold out of five different pre-defined dataset splits. For \textsc{Istella} and \textsc{Yahoo} there is only a single pre-defined fold, so we re-use the same train/validation/test split for each seed. The implicit labels generated per random seed and dataset are the same for each method evaluation. Overall statistics on the datasets are given in Tab.~\ref{tab:stats}. The distributions of these simulated labels are given in Fig.~\ref{fig:labels}. They largely correspond with statistics on internal datasets at Amazon Search.

For \textsc{WEB30k} and \textsc{Istella}, we use a quantile transform, trained on each training data split, with a normal output distribution. The \textsc{Yahoo} data already provides quantile features, so we just apply a standard transformation to also make them normally distributed.

\subsubsection{Methods}
In Sec.\ref{sec:rankformer}, we proposed the RankFormer architecture. To validate the benefit of the listwide loss, we evaluate the RankFormer for $\alpha \in \{0, 0.25, 1\}$, where $\alpha = 0$ implies that no listwide loss is used. The architecture then mostly reverts to the listwise Transformer discussed in Sec.~\ref{sec:transformer}, which has been evaluated using explicit labels in prior work \cite{pobrotynContextAwareLearningRank2021, qinARENEURALRANKERS2021,pangSetRankLearningPermutationInvariant2020}.

We evaluate the benefit of listwise attention by comparing these methods to a simple \textit{Multi-Layer Perceptron} (MLP) as a baseline. Moreover, as it is known that neural LTR methods often struggle to compete with tree-based models \cite{qinARENEURALRANKERS2021}, we train a \textit{Gradient-Boosted Decision Tree} (GBDT) \cite{friedmanGreedyFunctionApproximation2001} using the LightGBM \cite{keLightGBMHighlyEfficient2017} framework with the LambdaRank \cite{burgesRanknetLambdarankLambdamart2010} objective. We noticed a substantial benefit from using GBDTs with more trees than previous benchmarks. Therefore, we report results with both 1,000 (as in \cite{qinARENEURALRANKERS2021}) and 10,000 trees.

All methods optimize an LTR loss (Softmax for the MLP and RankFormer, LambdaRank \cite{burgesRanknetLambdarankLambdamart2010} for the GBDT) over the implicit labels for all training set lists $I$ where $t_I > 0$ (i.e. $\exists i \in I: y_i > 0$). Only the RankFormer with $\alpha > 0$ also trains on lists with $t_I = 0$. A hyperparameter search was done for the most important hyperparameters of each method, with the validation $\ndcgy$ as the objective. We refer to Appendix~\ref{app:hyperparams} for further details.

\subsubsection{Metrics}
The test set predictions in each experiment run are evaluated with the NDCG@10 metric, multiplied by 100. In our results, we report both $\ndcgy$, i.e. the NDCG with respect to the simulated \textit{implicit} labels $y_i$ (see Sec.~\ref{sec:simulating}), and the $\ndcgr$, i.e. the NDCG with respect to the original \textit{explicit} labels $r_i$. Lists with constant label vectors $\by_I$ or $\br_I$ respectively are not taken into account for the mean aggregation of the implicit or explicit NDCG.

Importantly, these metrics are only computed on the \textit{sampled} lists according to Sec.~\ref{sec:select}. This means that the NDCG computed on explicit labels is not comparable to other benchmarks, where the full selection (with better and worse items) is used instead. Note that the scores would anyhow not be comparable, since all methods are trained on simulated, implicit labels. To anyhow validate our experiment pipeline, we also report experiment results on the original \textsc{WEB30k} dataset in Appendix~\ref{app:original_results}, where the same hyperparameters were used as for the simulated experiments. 

\subsection{Results}\label{sec:results}
We report our results in Tab.~\ref{tab:main}. They largely confirm the common finding that strong GBDTs are superior to neural methods (e.g. the MLP and RankFormer) on tabular data \cite{grinsztajnWhyTreebasedModels2022}. Indeed, the strongest neural ranking methods already struggle to outperform a GBDT with 1,000 trees in recent benchmarks \cite{pangSetRankLearningPermutationInvariant2020,qinARENEURALRANKERS2021}. We now allow GBDTs an even higher complexity (i.e. 10,000 trees) and find that the advantage of GBDTs is even more distinctive: a GBDT model scores best on two (\textsc{Istella} and \textsc{Yahoo}) out of three datasets.

Yet despite the fact that GBDTs remain state-of-the-art on purely tabular data, neural methods may still hold more potential, e.g. because they integrate well with other data formats such as raw text and images. When only considering the neural methods, we see that the RankFormer with $\alpha = 0.25$ clearly benefits from learning with listwide labels. This confirms our hypothesis that there is value in learning from lists with no labels. However, the choice of $\alpha$ is important, as a larger focus on the listwide loss leads to a smaller focus on the listwise LTR objective. This can ultimately lead to worse NDCG scores (e.g. for $\alpha = 1$).

Finally, we note that there is no advantage for the RankFormer over the MLP baseline on the \textsc{Yahoo} dataset. The fact that Transformer-based methods appear to perform poorly on \textsc{Yahoo} is also aligned with previous benchmarks \cite{pangSetRankLearningPermutationInvariant2020, qinARENEURALRANKERS2021}.

Overall, we conclude that the RankFormer is superior to the state-of-the-art in neural LTR when listwide labels are provided. This advantage is seen both in the NDCG as measured on the simulated, implicit labels and the original, explicit labels.

\section{Experiments on Amazon Search}\label{sec:internal}
Much of the motivation for our proposed RankFormer method and our simulation of implicit labels has come from the application of LTR to e-commerce product search. In this section, we report results on Amazon Search data, which is collected from users interacting with Amazon websites. This results in implicit feedback, e.g. \textit{clicks} and \textit{purchases}.

We run both \textit{offline} and \textit{online} experiments.

\subsection{Offline Experiments}
\begin{table}[htb]
    \caption{$\ndcgy$@10 on an e-commerce dataset at \textsc{Amazon Search}, computed over organic \textit{implicit} labels. We report the overall $\ndcgy$ and the $\ndcgy$ specifically for queries that resulted in a purchase (i.e. $t_I = 2$).}
    \label{tab:offline}
    \begin{tabular}{ll||cc}
        & & \multicolumn{2}{c}{\textsc{Amazon Search} (offline)} \\ 
        & & $\ndcgy$ & $\ndcgy (t_I = 2)$ \\ \hline \hline
        \multirow{2}{*}{GBDT} & $\text{\# trees} = 10^3$ & $35.11$ & $50.00$\\
         & $\text{\# trees} = 10^4$ & $37.07$ & $53.70$\\ \hline
        MLP &  & $36.54$ & $53.98$\\ \hline
        \multirow{3}{*}{RankFormer} & $\alpha = 0$ & $37.05$ & $54.26$\\
         & $\alpha = 0.25$ & $\bm{37.33}$ & $\bm{55.20}$\\
         & $\alpha = 1$ & $37.18$ & $54.36$\\
        \end{tabular}
    \end{table}

The training set consists of 500k queries sampled from a month of customer interactions. Additionally, we gathered 139k validation set queries and 294k test set queries, both collected over one week. Between each set, there is a span of two weeks where no data is used to prevent leakage. In contrast to the public dataset experiments, we use organically collected implicit values without simulation. Our evaluation setup was similar to Sec.~\ref{sec:eval}, though no explicit labels are available for evaluation.

We report the results in Tab.~\ref{tab:offline}. Here, the RankFormer model clearly outperforms the baseline methods, especially on the $\ndcgy$ measured over queries that resulted in a purchase (i.e. $t_I = 2$). It sees a further benefit for $\alpha > 0$, indicating that the RankFormer is successful in learning from listwide labels. Since many queries have no individual product interactions ($\by_I = 0$), listwide labels can indeed be a useful signal.  

Compared to experiments on public datasets in Tab.~\ref{tab:main}, it is noticeable that the neural methods outperform the most powerful GBDTs. In fact, a GBDT with 10,000 trees is far too complex to meet the latency constraints required for online inference. 
An important factor contributing to this superiority of neural methods is that a richer set of features is available in the \textsc{Amazon Search} data than in the public dataset. For example, neural models can directly learn from raw text and image features, or from dense feature vectors generated by upstream neural models.

\subsection{Online Experiments}
The superiority of the RankFormer method in offline experiments may not translate to an advantage in practice. This can only be fully validated through online experiments. Yet, though the RankFormer benefits from the high parallelizability of the Transformer, its implementation demands significant changes to our internal pipeline, in particular because it performs listwise scoring as opposed to the pointwise scoring done by the GBDT and MLP. Therefore, we employ \textit{knowledge distillation} \cite{gouKnowledgeDistillationSurvey2021} by training a GBDT with 2,000 trees on the scores given by either the MLP or RankFormer (with $\alpha = 0.25$) that performed best in Tab.~\ref{tab:offline} as teacher models.

These two models (i.e. the GBDT with the MLP teacher and the GBDT with the RankFormer teacher) were then compared in an interleaving experiment as described in \cite{biDebiasedBalancedInterleaving2022}. Interleaving has a much greater statistical power than A/B tests, because the same customer is presented with a mix of the rankings of both models. The comparison between methods is quantified as the \textit{credit} lift, i.e. the extent to which the user's actions are attributed to either model. It was empirically shown in \cite{biDebiasedBalancedInterleaving2022} that this interleaving credit is strongly correlated with corresponding A/B metrics.

Our interleaving experiment ran for two weeks. Overall, the GBDT with the RankFormer teacher was credited with $13.7\%$ lift in revenue attributed to product search. The actual gain that would be seen in an A/B test is likely far smaller, yet the experiment shows that the GBDT with the RankFormer teacher is superior with $\bm{p = 0.02}$. Thus, we conclude that the listwise ranking performed by the RankFormer can be successfully translated to practical results, even when only distilling its listwise ranking scores to a pointwise GBDT scoring function. Though pointwise models lack the full list context, we hypothesize that powerful listwise rankers can help `fill in' implicit labels that are missing or noisy in the original data. Moreover, listwise models may favor features that lead to more diverse rankings. 

\section{Conclusion}
We formalized the problem of learning from listwide labels in the context of listwise LTR as a method to learn from both relative user feedback on individual items and the absolute user feedback on the overall list. Our proposed RankFormer architecture combines both objectives by jointly modelling both the overall quality of a ranking list and the individual utility of its items.

To conduct experiments on popular, public datasets, we simulated implicit feedback derived from the explicit feedback provided in those datasets. These experiments indicate that the RankFormer is indeed able to learn from listwide labels in a way that helps performance in LTR, thereby improving upon the state-of-the-art in neural ranking. However, in this tabular data setting, our results also show that strong GBDT models can be too powerful to be outclassed by neural methods. 

On internal data at Amazon Search, where a richer set of features is available, the RankFormer achieves better performance offline than any of the other methods, including strong GBDT models. In an online experiment, we distill the RankFormer to a practically viable GBDT and observe that this advantage is maintained over a complex MLP distilled to a similar GBDT.

Our findings encourage future work in learning from listwide labels. The user's perspective is central to ranking problems, and this particular signal of user feedback appears ripe for further research.

\begin{acks}
This research was supported by the TEN Search team at Amazon, the ERC under the EU's 7th Framework and H2020 Programmes (ERC Grant Agreement no. 615517 and 963924), the Flemish Government (AI Research Program), the BOF of Ghent University (PhD scholarship BOF20/DOC/144), and the FWO (project no. G0F9816N, 3G042220).
\end{acks}

\bibliographystyle{ACM-Reference-Format}
\balance
\bibliography{references}

\clearpage
\appendix
\section{Results on the Original \textsc{WEB30k}}\label{app:original_results}

\begin{table}[htb]
\caption{NDCG@10 on the \textsc{WEB30k}, computed over the original, \textit{explicit} labels and \textit{without subsampled lists}. Each result is the mean out of five runs, $\pm$ the standard error.}
\label{tab:no_select}
\begin{tabular}{ll||cc}
& & \textsc{WEB30k} \\ 
& & $\ndcgr$ \\ \hline \hline
\multirow{2}{*}{GBDT} & $\text{\# trees} = 10^3$ & $47.42\pm0.12$\\
    & $\text{\# trees} = 10^4$ & $51.05\pm0.10$\\ \hline
MLP &  & $49.30\pm0.10$\\ \hline
\multirow{2}{*}{RankFormer} & $\alpha = 0$ & $\bm{52.24\pm0.08}$\\
    & $\alpha = 0.25$ & $49.75\pm0.16$\\
\end{tabular}
\end{table}

Our results on the public datasets reported in Tab.~\ref{tab:main} were on the subsampled lists. This subsampling step was necessary to simulate realistic implicit feedback. To make these models more comparable to related work, we therefore run all our methods again on the original \textsc{WEB30k} without any of the simulation steps described in Sec.~\ref{sec:simulating}, yet with the same hyperparameters (as in Appendix~\ref{app:hyperparams}). These results are reported in Tab.~\ref{tab:no_select}.


In contrast to our experiments on simulated data, we now observe that the RankFormer with $\alpha = 0.25$ achieves far worse results than with $\alpha =0$. The listwide loss may be less informative here, because the explicit labels were assigned independently per item by human annotators. The listwide quality effect is then far less important, which means smaller $\alpha$ values should be used.

\section{Architecture Details of the listwise Transformer}\label{app:transformer_details}
The Transformer (TF) architecture that was used as a basis for the RankFormer, was implemented using the TransformerEncoderLayer\footnote{https://pytorch.org/docs/1.13/generated/torch.nn.TransformerEncoderLayer.html} of the PyTorch framework. This implementation has an output dimensionality $d_\text{val}$ and an attention dimensionality $d_\text{att}$ that equals the input dimensionality $d_\text{x} = d_\text{att} = d_\text{val}$. It uses a LayerNorm\cite{baLayerNormalization2016} (LN) transformation that is applied on the \textit{input} of both the Self-Attention (SA) and Feed-Forward (FF) blocks in our configuration. Moreover, it uses residual connections for each block. The output of a single TransformerEncoderLayer $\TF^{l}$ is thus given by 
\begin{equation}
\TF^l(\bX_I) = \bX_I + \widetilde{\text{FF}}\left(\bX_I + \widetilde{\SA}\left(\bX_I\right)\right)
\end{equation}
with $\widetilde{\SA}(\cdot) = \SA(\text{LN}(\cdot))$ and $\widetilde{\text{FF}}(\cdot) = \text{FF}(\text{LN}(\cdot))$. The SA and FF blocks also contain Dropout layers and the FF uses a GELU activation. The Transformer $\TF$ is then simply a composition of $N_l$ Transformer layers: $\TF = \TF^{N_l}\left(...\left(\TF^1(\bX_I)\right)\right)$.

Typically, the parameters of $\TF$ are optimized by computing the loss for a batch of lists. If the lists in a batch do not share the same length
then all lists are concatenated with feature vectors of zeros $\mathbf{0} = (0)_{i=0}^{d_x}$ until all lists have the same length as the longest list in the batch.

\begin{table*}[ht]
    \caption{Hyperparameter selections and ranges for the public dataset experiments. For GBDT, the HPO was later performed separately with $10^3$ and $10^4$ trees.}
    \label{tab:hyper}
    \begin{tabular}{cl||c|c}
     & & Best & Range \\ \hline\hline\vrule width 0pt height 2.7ex
    \multirow{5}{*}{GBDT} & \texttt{num\_iterations} & \begin{tabular}{@{}ll@{}} $10^3$ & $10^4$ \\\end{tabular} & $[10^2, 10^4]$\\
     & \texttt{min\_data\_in\_leaf} & \begin{tabular}{@{}ll@{}} $255$ & $376$ \\\end{tabular} & $[10, 500]$\\
     & \texttt{num\_leaves} & \begin{tabular}{@{}ll@{}} $507$ & $271$ \\\end{tabular} & $[128, 512]$\\
     & \texttt{learning\_rate} & \begin{tabular}{@{}ll@{}} $0.46$ & $0.32$ \\\end{tabular} & $[0.01, 0.5]$\\
     & \texttt{cegb\_tradeoff} & \begin{tabular}{@{}ll@{}} $0.13$ & $1.65$ \\\end{tabular} & $[0.1, 10]$\\ \hline\vrule width 0pt height 2.7ex
    
    \multirow{4}{*}{MLP} & \texttt{lr} & $10^{-3}$ & $\{10^{-4}, 10^{-3}, 10^{-2}\}$\\
    & \texttt{weight\_decay} & $10^{-1}$ & $\{0, 10^{-2}, 10^{-1},1\}$\\
    & \texttt{dropout} & $0.25$ & $\{0.1, 0.25, 0.4, 0.5\}$\\
    & \texttt{hidden\_dims} & $[512,256,128]$ & $[1,5]$ layers with size $[128, 1024]$\\ \hline
    
    \multirow{4}{*}{RankFormer} & \texttt{nhead} & $1$ & $\{1, 2, 4\}$\\
    & \texttt{dim\_feedforward} & $512$ & $\{128, 256, 512\}$\\
    & \texttt{dropout} & $0.25$ & $\{0.1, 0.25, 0.4, 0.5\}$\\
    & $N_l$ & $3$ & $\{2, 3, 4\}$\\
    \end{tabular}
    \end{table*}

\section{Label Simulation Details}
In Sec.~\ref{sec:simulating}, we discuss our simulation of implicit labels on public datasets in three stages: selection, intent and interaction. To ensure reproducibility, we write out the pseudocode of the complete process in Alg.~\ref{alg:simulation}.

Note that for the intent $T_I$, we directly sample from its distribution as given by Eq.~\ref{eq:intent}. However, the actual implicit label $y_i$ is sampled according to a cascade of probability distributions defined in Eq.~\ref{eq:interaction}. 


\begin{algorithm}[h]
\SetAlgoLined
\DontPrintSemicolon
\KwData{Explicit labels $r_i \in \{1, ..., \rmax\}$ for item $i \in J$ for a superset of original lists $J \in \mathcal{J}$}
\KwParameters{maximum list length $N_s$\qquad\qquad bootstrapping amount $N_b$\qquad\qquad\qquad
relevance grade map $\rho: \{1, ..., \rmax\} \to [0, 1]$ \qquad\qquad
conversion frequency $\kappa$\qquad\qquad\qquad\qquad
click noise $\epsilon$}
\KwResult{Implicit labels $y_i \in \{0, 1, 2\}$ for item $i \in I$ for a superset of sampled lists $I \in \mathcal{I}$.}
$\mathcal{I} \gets \emptyset$\;
\For{$J \in \mathcal{J}$}{
    \tcc{1. Selection}
    \For{$n = 1$ \KwTo $N_b$}{
        \If{$\abs{J} > N_s$}{
            $I \gets \emptyset$\;
            \While{$\abs{I} < N_s$}{
                $I \gets I \cup \unif(J)$\;
            }
        }
        \Else{
            $I \gets J$\;
        }
    }\;
    \tcc{2. Intent}
    $t \gets \unif([0, 1])$\;
    $r \gets \max_{i \in I} r_i$\;
    $T_I \gets \begin{cases}
        0 & t < 1 - \rho(r)\\
        1 & t \in [1 - \rho(r), (1 - \kappa)\rho(r)]\\
        2 & t > (1 - \kappa)\rho(r)
    \end{cases}$
    \;\;

    \tcc{3. Interaction}
    \For{$i \in I$}{
        \If{$T_I = 2$ \KwAnd $\rho(r_i) > \unif([0, 1])$}{
            $y_i \gets 2$
        }
        \ElseIf{$T_I \geq 1$ \KwAnd $\epsilon + (1 - \epsilon) \rho(r_i) > \unif([0, 1])$}{
            $y_i \gets 1$
        }
        \Else{
            $y_i \gets 0$
        }
    }
$\mathcal{I} \gets \mathcal{I} \cup I$\;
}
\caption{Implicit label simualtion as described in Sec.~\ref{sec:simulating}. The $\unif(S)$ operator randomly returns a value sampled from set $S$ with a uniform probability distribution.}
\label{alg:simulation}
\end{algorithm}

\section{Hyperparameters}\label{app:hyperparams}
Our hyperparameter optimizations (HPOs) were performed separately for the public dataset experiments in Sec.~\ref{sec:public} and the Amazon Search experiments in Sec.~\ref{sec:internal}. For the public datasets, we only performed HPOs on Fold 1 of the \textsc{WEB30k} dataset with simulated implicit labels. The hyperparameters with the best $\ndcgy$ on the validation set were then used for the other Folds and for the \textsc{Istella} and \textsc{Yahoo} datasets. We report all optimized values and their ranges in Tab.~\ref{tab:hyper}.

As the hyperparameter space is relatively limited for GBDTs, we conducted a full Bayesian Optimization there, separately for 1,000 and 10,000 trees. We use LightGBM \cite{keLightGBMHighlyEfficient2017} to implement the GBDTs, so we directly list the parameter names as they occur in that framework.

The convergence parameters were first optimized for the MLP, after which they were also used for the RankFormer. In total, we train for 200 epochs and used the Adam optimizer with parameters \texttt{lr} and \texttt{weight\_decay} that starts decaying with an inverse square root schedule after 20 epochs. Additionally, we used \texttt{dropout} for the MLP and hidden layer sizes \texttt{hidden\_dims}.

For the RankFormer, we used the convergence and setting of the MLP. For each layer of its Transformer $\TF^l$ (discussed in Appendix~\ref{app:transformer_details}), the \texttt{nhead}, \texttt{dim\_feedforward} and \texttt{dropout} were optimized in the PyTorch implementation. The full Transformer component consists of $N_l$ of such layers. To reduce further complexity, we directly choose the $h_s$ and $\bh_d$ scoring heads to be feed-forward networks with a single 128-dimensional layer.
\end{document}